\documentclass{article}
\usepackage{natbib}
\usepackage{bm}
\usepackage{amsfonts,amssymb,amsmath,amsbsy,amscd,latexsym}
\usepackage{graphicx}
\usepackage{enumerate}
\usepackage{epstopdf} 
\usepackage{xcolor}
\usepackage[breaklinks=true]{hyperref}
\hypersetup{
allcolors       = {blue},
linkbordercolor = {white},
linkcolor       = {cyan},
colorlinks      = true
}
\renewcommand{\Im}{\mathop{\mathrm{Im}}} 
\renewcommand{\Re}{\mathop{\mathrm{Re}}} 

\newcommand{\dd}{\mathrm{d}}                        
\newcommand{\ii}{\mathrm{i}}                        
\newcommand{\bxi}{{\bm{\xi}}}              
\newcommand{\bb}{\mathbf{b}}             
\newcommand{\be}{\mathbf{e}}             
\newcommand{\bk}{\mathbf{k}}             
\newcommand{\bbm}{\mathbf{m}}            
\newcommand{\bn}{\mathbf{n}}             
\newcommand{\br}{\mathbf{r}}             
\newcommand{\bs}{\mathbf{s}}             
\newcommand{\bV}{\mathbf{V}}             
\newcommand{\cL}{c_{\text{L}}}

\newcommand{\cS}{c_{\text{S}}}
\newcommand{\cSL}{c_{\text{S,L}}}

\newcommand{\vL}{v_{\text{L}}}           
\newcommand{\vS}{v_{\text{S}}}           
\newcommand{\vSL}{v_{\text{S,L}}}        
\newcommand{\calL}{{\mathcal{L}}}

\newcommand{\red}[1]{{\color{red} #1}}

\renewcommand{\vS}{\overline{v}_{\text{S}}}
\renewcommand{\vL}{\overline{v}_{\text{L}}}
\newcommand{\vP}{\overline{v}_{\text{P}}}
\newcommand{\cP}{c_{\text{P}}}

\renewcommand{\vSL}{\overline{v}_{\text{S},\text{L}}}

\newcommand{\zS}{z_{\text{S}}}
\newcommand{\zL}{z_{\text{L}}}
\newcommand{\wL}{\widetilde{L}}
\begin{document}
\title{{\bf Dynamic Peach-Koehler self-force, inertia, and radiation damping of a regularized dislocation\quad V.3}}
\author{
Yves-Patrick Pellegrini~$^\text{a,b,}$\footnote{{\it E-mail address:} yves-patrick.pellegrini@cea.fr},
\\ \\
${}^\text{a}$ CEA, DAM, DIF, F-91297 Arpajon, France.\\
\\
${}^\text{b}$ Universit{\'e} Paris-Saclay, CEA,\\ Laboratoire Mati{\`e}re sous Conditions Extr{\^e}mes,\\ 91680 Bruy{\`e}res-le-Ch{\^a}tel, France.\\
}
\date{\today}
\maketitle
Dislocations; Dynamic Peach-Koehler force; Regularized fields; Radiation damping; Inertia; Supersonic motion.
\begin{abstract}
The elastodynamic Peach-Koehler force is computed for a fully-regularized straight dislocation with isotropic core in continuum isotropic elastic elasticity, in compact forms involving partial mass or impulsion functions relative to shear and compressional waves. The force accounts for both dynamic radiation damping and inertia. The expressions are valid indifferently for subsonic or supersonic velocities. Results are compared with the case of a flat-core dislocation of the Peierls-Eshelby type, for a motion of jump from rest to constant velocity. In the steady-state limit, the Lagrangian function relevant to expressing the force in the flat-core case must be replaced by a related but different function for the regularized dislocation. However, by suitably defining the regularizing dislocation width, the steady-state limits of the force for the fully-regularized and flat-core dislocations can be matched exactly.
\medskip

\red{This Version 3 corrects several typographical errors, highlighted in red in the text: (i) in Eqs.\ (52a), (52b), and in Eqs.\ (53a), (53b) (inversion of exponents); (ii) in the caption of Fig.\ 1 (extraneous value); and (iii) in the rightmost expression of Eq.\ (61) (final equality). All other results remain unchanged. In particular, the plots in Fig. 1 were produced using the correct expressions and are therefore unaffected. The author is grateful to Kyriann Loué (student ENPC) for identifying these errors; any remaining errors are solely the author's responsibility.}
\end{abstract}

\section{Introduction}
Dislocations are linear defects whose motion is responsible for the plastic deformation of metals \citep{ANDE17}. A given dislocation moves under the cumulative action of the externally applied stress, of the stress generated by other dislocations, and of its self-stress. All these contributions are accounted for by the Peach-Koehler (PK) force \citep{PEAC50}. The present work is concerned with the PK self-force, in dynamic situations where the retarded character of the elastodynamic interactions is of importance. In connection with elastodynamics the current state of knowledge of the mechanics and physics of high-speed dislocations has been reviewed by \citet{GURR20}. The dynamic PK force must be evaluated to devise consistent elastodynamic Discrete Dislocation codes \citep{GURR15a,CUIP19}. Although it plays a major role in the dynamic mobility law of dislocations, its influence in the long-term plastic response of metals \citep{GURR16} remains to be further assessed, especially with regard to individual interaction processes between dislocations at high stain rate \citep{PILL06}. From a technical standpoint, it is well-defined (in the sense that it takes on definite values everywhere) only for a dislocation of finite width: the elastodynamic stress field of a Volterra dislocation (of zero-width core) is a mathematical distribution, or generalized function. As such, if interpreted as a usual function, it possesses wavefront singularities \citep{MARK82} that need to be tamed by an appropriate \emph{regularization scheme} able to account for the finite width of the physical dislocation. A number of regularization schemes are available \citep{CAIA06,LAZA09,POLA18}. In this context, the dislocation density acts as a test function for the fields. The present work is based on the scheme introduced by \citet{PELL15}, that proved very convenient to addressing elastodynamic dislocation problems \citep{LAZA16}.

The nonzero dislocation width results from geometric nonlinearities in elastic field theory \citep{ACHA01}, and from nonlinearity in the interatomic forces \citep{HUNT14b}. For this reason, recourse to an atomistic approach to the self-force \citep{CHO17} has been advocated \citep{LUBA19} in order to achieve quantitative agreement with atomistic simulations. However, since inter-atomic forces govern the core shape (possibly time-varying), specifying empirically this shape by regularization within the framework of linear elasticity can in principle deal with this issue at the crudest approximation order. In particular, such an approach allows one to approach in general terms the mathematical structure of the dynamic PK (DPK) self-force.

This DPK force accounts for both inertia and radiation damping (or radiative drag) of dislocations, which are two sides of the same coin \citep{NABA51b,PELL12}. Pioneering works have long been confined to the `nonrelativistic regime' of dislocation velocities small with respect to the wavespeeds in the medium. Only recently have non-singular expressions able to cope with any velocity been proposed \citep{PELL14}. In the continuum, radiation damping takes place either in accelerated/decelerated motion by emission of acceleration or braking waves, or in steady motion at supersonic velocities, in the form of energy losses via Mach cones \citep{LAZA16}.\footnote{In the enlarged context of theories of systems of particles coupled to fields, in which waves are radiated, the problem of the self-force has deep and still much debated implications with regard to causality; e.g., \citep{ROVE04,BOOZ13}.} Formation of Mach cones in elasticity is sometimes referred to as the \emph{elastic Cerenkov effect} \citep{BERC04,LICA20}.

If the lattice structure of the medium is considered, dislocations are moreover subjected to specific effects. The (conceptually) simplest one is due to the dispersion relations of the crystal, which notably differ from the linear dispersion relations of continuum elasticity \citep{ATKI65}. This leads to anisotropic response, and modifies the notion of supersonic velocity \citep{ESHE56}. However, no complete and undisputable theory of lattice dislocation dynamics at very high velocity is available so far. Moreover, specific lattice-induced dissipative processes take place \citep{ALSH75,NADG88}, among which the so-called phonon-wind effect. It involves a non-linear coupling between the dislocation strain field and the phonon field. It has recently been computed in the subsonic regime within the Debye approximation and isotropic elastic response, leading to a supplementary effective drag force on the dislocation, $f^{\rm drag}=-B(v) v$, where the drag coefficient $B(v)$ depends on velocity and on the dislocation character \citep{BLAS19,BLAS20}. However, this function $B(v)$ is still unknown in the supersonic regime. Such lattice-induced processes are left out from the present study, focused exclusively on radiation damping and inertia in continuum mechanics, which can be computed for any velocity.

Accordingly, the purpose of this work is to investigate, within isotropic linear elasticity, the influence of the regularization scheme on the elastodynamic self-force for straight dislocations in an infinite medium (for interface effects, see, e.g., \citep{GURR15b}). To this aim the DPK self-force consistent with the work of \citet{PELL15} is evaluated in terms of effective mass functions \citep{KOSE79,HIRT98,NIMA08,PELL12,PELL14}. The result is intended for use as a control for the force, for limiting straight dislocations, in a three-dimensional elastodynamic discrete--dislocation simulation method under development, based on the 3D regularizing function \eqref{eq:delta3} below (Pellegrini \emph{et al.}, in preparation). For completeness, the three dislocations characters are considered, including the possibility of a `climb'-edge component normal to the glide plane that would move at high velocity along with a `glide'-edge component. Admittedly speculative, this possibility \citep{WEER67a,WEER67b},\footnote{See also \citep{PELL10}.} cannot be excluded in the context of misfit dislocations (disconnections) gliding on a habit plane, reviewed by \citet{POND03}.\footnote{The two media separated by the plane would then need to be assumed (rather academically) of identical isotropic elastic properties.}

The present work will make clear that, irrespective of the regularization considered, the in-plane component of the DPK  force (per unit dislocation length) reduces in the Volterra limit to an ill-defined formal expression of the \emph{`mass-form'} type
\begin{subequations}
\begin{align}
\label{eq:template}
f^{\text{PK}}(t)&=-2\int_{-\infty}^{t} \frac{\dd t'}{t-t'}m(\overline{v})\frac{\dd\overline{v}}{\dd t'}+(\text{undetermined}).
\end{align}
where, with $\xi(t)$ the position of the dislocation,
\begin{align}
\label{eq:meanveltemplate}
\overline{v}(t,t')&=\frac{\xi(t)-\xi(t')}{t-t'},
\end{align}
\end{subequations}
and $m(v)$ is the \emph{prelogarithmic mass function} relative to the dislocation character considered (see next section). Equation \eqref{eq:template} is mathematically meaningless: the integral is logarithmically divergent (infinite) at $t'=t$, and can include another infinite term, denoted as `undetermined'; see Sec.\ \ref{sec:flatcore} for an example. It nonetheless provides a template to which any admissible DPK-force expression should be amenable in the limit of vanishing core size, irrespective of the regularization procedure employed. The mean velocity \eqref{eq:meanveltemplate} between instants $t$ and $t'$ appears naturally in this self-interaction problem \citep{ESHE53,PILL07} and, more generally, stands as the hallmark of a collective-variable treatment when a moving defect self-interacts via retarded interactions such as, e.g., a moving magnetic domain wall of finite width \citep{BOUC90}. In a collective-variable treatment, the defect shape and position are hypothesized to evolve through a small number of time-dependent variables. In expression \eqref{eq:template} only the position variable is considered. Recognition of the fundamental structure of Eq.\ \eqref{eq:template} \citep{PELL12} followed earlier attempts such as the one by \citet[Eq.\ (7-40)]{ANDE17} for the screw dislocation, or by \citet[Eq.\ (11)]{PILL07} for screw and edge dislocations, which involved the instantaneous velocity rather than the mean one.

The next sections illustrate how different regularization schemes turn \eqref{eq:template} into a useful expression. Energy-related functions, as computed for a Volterra dislocation, are reviewed in the Sec.\ \ref{sec:erfuncts}. Section \ref{sec:flatcore} recalls the peculiarities of the DPK self-force for a semi-regularized (flat-core) Peierls-Eshelby dislocation. Our regularization scheme is introduced Sec.\ \ref{sec:dynPK}, which offers a general approach to computing the force for a regularized loop and for a straight dislocation. `Mass' and ìmpulsion' forms of the force are derived in Sec.\ \ref{sec:massimp}. Initial conditions and a few particular motions are discussed in Sec.\ \ref{sec:initcond}, which presents numerical results and comparisons between the flat-core and the regularized dislocations. Section \ref{sec:concl} concludes the paper.

\section{Energy-related functions}
\label{sec:erfuncts}
Consider a straight Volterra dislocation of Burgers vector $\bb$ that moves at arbitrary velocity steady $v$ in an isotropic medium. All energy-related functions, such as field energy, Lagrangian, impulsion, and mass, can be expressed as two-dimensional lo\-ga\-rith\-mi\-cal\-ly-divergent integrals of appropriate fields over the plane transverse to the dislocation line. These integrals must be tamed by specifying an upper cut-off $R$ (the system size) and a lower one $r_0$ (of order the dislocation core size) on the integration radius. A factor $\log(R/r_0)$ results for all these functions \citep{HIRT98}, and only the multiplicand -- the so-called \emph{logarithmic prefactor} -- is of interest. Thus, when referring hereafter to energy-related functions, we shall in fact always refer by a slight abuse of language to their logarithmic prefactor.

The Lagrangian function $L(v)$ is defined as the difference between the kinetic and elastic energy \emph{of the field} of the steadily-moving dislocation. The medium has Lam\'e elastic constants $\lambda$ and $\mu$, density $\rho$, and shear and longitudinal wavespeeds $\smash{\cS=\sqrt{\mu/\rho}}$ and $\smash{\cL=\sqrt{(\lambda+2\mu)/\rho}}$, respectively. Lagrangian functions read, for the screw (s), glide-edge (g) and climb-edge (c) dislocation characters \citep{WEER67a,BELT68},
\begin{subequations}
\begin{align}
\label{eq:lags}
L^{\rm s}(v)&=-w_0\,\sqrt{1-v^2/\cS^2},\\
L^{\rm g}(v)&=-w_0\,4\left(\frac{\cS}{v}\right)^2\left\{\sqrt{1-v^2/\cL^2}-\frac{[1-v^2/(2\cS^2)]^2}{\sqrt{1-v^2/\cS^2}}\right\},\\
\label{eq:lagc}
L^{\rm c}(v)&=-w_0\,4\left(\frac{\cS}{v}\right)^2\left\{\sqrt{1-v^2/\cS^2}-\frac{[1-v^2/(2\cS^2)]^2}{\sqrt{1-v^2/\cL^2}}\right\},
\end{align}
\end{subequations}
where, with $b$ the modulus of the Burgers vector, $w_0=\mu b^2/(4\pi)$ is the reference line energy density per unit dislocation length. From these expressions, the impulsion $p(v)$ and mass $m(v)$ functions are obtained as the derivatives $p(v)=L'(v)$ and $m(v)=L''(v)$.

Introducing the wavespeed ratio
\begin{align}
\phi&=\cL/\cS,
\end{align}
the impulsion functions deduced from \eqref{eq:lags}--\eqref{eq:lagc} are
\begin{subequations}
\begin{align}
\label{eq:ps}
p^{\rm s}(v)&=\frac{w_0}{\cS}\frac{v}{\cS}\sqrt{1-v^2/\cS^2},\\
\label{eq:pg}
p^{\rm g}(v)&=\frac{w_0}{\cS}\left(\frac{\cS}{v}\right)^3
\left[4\left(1-\frac{v^2}{\cL^2}\right)^{-1/2}\left(2-\frac{v^2}{\cL^2}\right)
-\left(1-\frac{v^2}{\cS^2}\right)^{-3/2}
\left(8-12\frac{v^2}{\cS^2}+2\frac{v^4}{\cS^4}+\frac{v^6}{\cS^6}\right)\right],\\
\label{eq:pc}
p^{\rm c}(v)&=\frac{w_0}{\cS}\left(\frac{\cS}{v}\right)^3
\left\{4\left(1-\frac{v^2}{\cS^2}\right)^{-1/2}\left(2-\frac{v^2}{\cS^2}\right)
-\left(1-\frac{v^2}{\cL^2}\right)^{-3/2}
\left[8-12\frac{v^2}{\cL^2}+2\phi^2\left(2-\phi^2\right)\frac{v^4}{\cL^4}+\phi^4\frac{v^6}{\cL^6}\right]\right\}.
\end{align}
\end{subequations}
Furthermore, the mass functions can be written as
\begin{subequations}
\begin{align}
\label{eq:ms}
m^{\rm s}(v)&=\frac{w_0}{\cS^2}\left(1-v^2/\cS^2\right)^{-3/2},\\
\label{eq:mg}
m^{\rm g}(v)&=\frac{w_0}{\cS^2}
\left(\frac{\cS}{v}\right)^4\left[
\left(1-\frac{v^2}{\cS^2}\right)^{-5/2}\left(24-60\frac{v^2}{\cS^2}+46\frac{v^4}{\cS^4}-7\frac{v^6}{\cS^6}\right)
-4\left(1-\frac{v^2}{\cL^2}\right)^{-3/2}\left(6-9\frac{v^2}{\cL^2}+2\frac{v^4}{\cL^4}\right)
\right],\\
\label{eq:mc}
m^{\rm c}(v)&=\frac{w_0}{\cS^2}
\left(\frac{\cS}{v}\right)^4\left\{
\left(1-\frac{v^2}{\cL^2}\right)^{-5/2}
\left[24-60\frac{v^2}{\cL^2}+2\left(24-2\phi^2+\phi^4\right)\frac{v^4}{\cL^4}
+\phi^2\left(\phi^2-8\right)\frac{v^6}{\cL^6}\right]\right.\nonumber\\
&{}\left.\hspace{2cm}-4\left(1-\frac{v^2}{\cS^2}\right)^{-3/2}\left(6-9\frac{v^2}{\cS^2}+2\frac{v^4}{\cS^4}\right)
\right\},
\end{align}
\end{subequations}
where the `screw' and `glide-edge' expressions have previously been given by \cite{HIRT98} in other equivalent forms. The above writing is more convenient for the present purpose.

The dislocation energy function is
\begin{align}
W(v)=v\,p(v)-L(v).
\end{align}
It is equal to the sum of the kinetic and elastic energies of the field, which both depend on $v$. Confusion sometimes arises in the literature between the kinetic energy \emph{of the material displacement field} and the kinetic energy \emph{of the dislocation}. The kinetic energy of the dislocation is
\begin{align}
\label{eq:kinenergy}
K(v)&=W(v)-W(0).
\end{align}
All of the above functions have a well-defined limit when $v\to 0$.

\section{Flat-core regularization: the Peierls(-Eshelby) dislocation}
\label{sec:flatcore}
Elastodynamic dislocation theory was initiated considering a flat-core dislocation of the Peierls type \citep{NABA51a,ESHE53}. This semi-regularized model has a finite-valued stress field everywhere, but the field is discontinuous at the slip plane since the solution can be obtained in the simplest case from a Volterra dislocation by ``cutting out the material between the planes $y=\pm a$ and gluing together the remainder'' \citep{ESHE49}.

There has been a long practice of dealing with subsonic and supersonic regimes of source motion by means of separate analytical treatments (e.g., \citet[Sec.\ 7.11]{ERIN75}). However, the `Peierls-Eshelby' model and some related ones are compatible with any speed (possibly supersonic) if the fields are formulated in terms of complex-valued times, and the self-force is formulated in terms of complex-valued speeds. Thus, compact analytical expressions that encompass all regimes are obtained \citep{PELL12,PELL14}. There are at least two reasons for this: (a) from the physical standpoint, and within the class of models considered, the possibility of a complex-valued velocity has been related \citep{PELL17} to the outgoing-wave radiation condition that implements causality at the fundamental level of the elastodynamic Green operator; (b) from the mathematical standpoint, field expressions for Volterra dislocations are generalized functions (distributions), and representations with complex numbers is a well-known method of regularizing such `functions'. Then, the occurrence of complex-valued velocities, of imaginary part determined by the core size, stands as a consequence of the simple analytical structure in the complex domain of the core-regularizing function used as a test function. These two aspects are interrelated because assessable wave emission is produced by a core of finite size.

Specifically, for the 'Peierls-Eshelby' model, the in-plane component of self-force has been derived in terms of the mass functions \eqref{eq:ms}--\eqref{eq:mg} as\footnote{In Ref.\ \citep{PELL14} this force, interpreted as an inertial force, is considered with the opposite overall sign.}
\begin{align}
\label{eq:selfforceflat}
f^{\text{PK, flat-core}}(t)&=-2\Re\int_{-\infty}^{t} \frac{\dd t'}{t-t'}m(\overline{v}(t,t'))\frac{\dd\overline{v}}{\dd t'}(t,t')-w_0\kappa\frac{2}{a(t)}\frac{\dot{\xi}(t)}{\cS},
\end{align}
where
\begin{align}
\label{eq:meanvelflat}
\overline{v}(t,t')&=\frac{\xi(t)-\xi(t')}{t-t'}+\frac{\ii}{2}\frac{a(t)+a(t')}{t-t'}
\end{align}
is a complex-valued velocity akin to a mean velocity between instants $t$ and $t'$ supplemented by an imaginary part that involves the time-dependent core width $a(t)$, $m(v)$ is the function appropriate to each dislocation character taken among Eqs.\ \eqref{eq:ms}--\eqref{eq:mc}, and $\kappa$ is a coefficient that depends on the dislocation character, to be explained next. In the Volterra limit, expression \eqref{eq:selfforceflat} falls back to the template \eqref{eq:template}, with the rightmost term leading to the additional `undetermined' infinite contribution.

In \eqref{eq:meanvelflat}, the imaginary part of the complex-valued mean velocity completely regularizes the equation of motion in which the self force intervenes, making its solution well-defined. It can be easily shown by the method employed to derive \eqref{eq:selfforceflat} that the form \eqref{eq:selfforceflat} holds as well for a `climb' component, with mass function \eqref{eq:mc}.

The following limits affect the expression \eqref{eq:selfforceflat} of the self-force. Assuming that in the remote past the dislocation is at rest or moves with constant speed, one has
\begin{subequations}
\label{eq:limitsflat}
\begin{align}
\lim_{t'\to t^-}\overline{v}(t,t')&=+\ii\,\infty,\\
\lim_{t'\to -\infty}\overline{v}(t,t')&=\dot{\xi}(-\infty)+\ii\,0^+.
\end{align}
\end{subequations}
The integral term in \eqref{eq:selfforceflat} represents retarded self-interactions as the dislocation moves on its slip plane, and fully accounts for inertia of elastodynamic origin. It is well-defined as $t'\to t^-$ because $m(+\ii\,\infty)=0$, see \eqref{eq:ms}--\eqref{eq:mc}. The rightmost term in \eqref{eq:selfforceflat} originates from the radiation damping term in the DPE caused by wave outflow from the slip plane. For screw or glide-edge dislocations, the constant $\kappa$ is equal to 1. However, an indirect argument has shown that its general expression in this model is $\kappa=(\cS/w_0)\Im[p(+\ii\,\infty)]$ \citep[Eq.\ (66)]{PELL12}, which is a consequence of the equal-time limit \eqref{eq:limitsflat}${}_1$. Hence, for a `climb-edge' Burgers vector component, $\kappa=\cL/\cS$.

Equation \eqref{eq:selfforceflat} was first derived \citep{PELL12} as part of a mean-field solution \citep{PELL14} of the dynamic Peierls integro-differential equation (DPE) \citep{PELL10}, with the denomination `mean-field' referring to the approximation procedure employed; namely, to reduce the numbers of degrees of freedom by imposing the master shape of the core, while allowing the core to vary only through the size parameter $a$ introduced as a scaling variable. In the DPE the core shape function is the unknown.

In this context, the radiative damping term has raised some debate \citep{MARK11,PELL11}. This issue is further documented here, by observing that the kernel for the stress on the slip plane in the DPE is the same as for a self-healing crack. In crack theory, the necessity for this damping term has long been known, which the present author was unaware of while rediscovering it independently in 2010. Apparently first hinted at  by \cite{RICE93} to account for the energy outflow of a seismic fault, this term was explicitly exhibited by \cite{COCH94} for a mode--III crack (which would correspond to a screw dislocation), and was subsequently further generalized to the other crack modes by \citet{PERR95}, \citet[Eqs.\ (25)--(26)]{GEUB95}, \citet{COCH97}; see also \citet{JOSI18}.\footnote{A typo in \citep[Eq.\ (26)]{GEUB95} has been corrected in \citep[Eq.\ (A.1)]{COCH97}, as pointed out in the later reference.} As the latter works make clear, calculations are most straightforward in the spectral domain, because the various possibilities of integrating by parts with respect to the space or time variables, within the boundary-integral expression for the dynamic stress, make the expression of the stress kernel in the spatio-temporal domain vary from one author to another. Of course, the $\kappa$ coefficient obtained from $p(v)$ as recalled above, matches its value obtained by other means in crack theory. Still, the relationship between the elastodynamic stress on the slip plane, and the Lagrangian and impulsion functions with a complex-valued velocity argument was derived in the dislocation context. While being undisputable, it is not yet fully understood and certainly deserves further investigations with regard to the physics of radiative wave emission.

For the flat-core dislocation, Equ.\ \eqref{eq:selfforceflat} determines $\xi(t)$ if the core is rigid, with $a(t)\equiv a$  constant. However, if the core is allowed to depend on time a second equation for $a(t)$ must be derived, on the basis of the energetics of the $\gamma$-surface in the Peierls model, as for the flat-core dislocation \citep{PELL14}. This point is not further considered hereafter.

\section{The dynamic Peach-Koehler self-force}
\label{sec:dynPK}
\subsection{Regularized dislocation density}
The Nye dislocation-density tensor is well-suited to representing dislocation cores, even at the atomistic level \citep{HART05b,WOOD08}. A regularized dislocation loop is represented by its Nye density tensor defined by the loop integral of line element $\dd\bs$
\begin{align}
\label{eq:loopdendyn}
\alpha_{ij}(\br,t)
&=b_i\oint_{\calL(t)}\delta^{(3)}_\varepsilon(\br-\bs)\dd s_j,
\end{align}
where $\delta_\varepsilon^{(3)}(\br)$ is the 3D regularizing function. It can be construed as a local spatial density of Burgers vector and characterized the rigid core shape of the dislocation. The present work uses the model
\begin{align}
\label{eq:delta3}
\delta_\varepsilon^{(3)}(\br)&=\frac{\varepsilon}{\pi^2(r^2+\varepsilon^2)^2}.
\end{align}
Integrating it along the $z$ coordinate yields the following 2D density
\begin{align}
\label{eq:delta2}
\delta_\varepsilon^{(2)}(\br)&=\frac{\varepsilon}{2\pi(r^2+\varepsilon^2)^{2/3}}.
\end{align}
The vector $\br=(x,y,z)$ in \eqref{eq:delta3}, whereas $\br=(x,y)$ in \eqref{eq:delta2}. In both cases, the 3D and 2D Fourier transforms of these functions is $\exp(-\varepsilon k)$, where $\bk$ stands for the 2D or 3D Fourier wavevector. Thus, these functions are of unit integral and reduce to the 3D or 2D Dirac distributions (case of a Volterra dislocation) when $\varepsilon\to 0$. A more involved core model, again exponential in the Fourier space but with additional directional dependence, has been considered by \citet{GUNT73}. Equation \eqref{eq:delta2} has been used to regularize the 2D elastodynamic problem of straight dislocations \citep{PELL15,LAZA16}. Compared to other popular regularizing functions used for dislocation problems \citep{CAIA06,POLA18}, the present ones benefit from the simpler autoconvolution property $\delta_\varepsilon\star\delta_\varepsilon=\delta_{2\varepsilon}$ in 2D or 3D, as the FTs immediately show (the star denotes convolution with respect to the space variable), which notably simplifies the computation of the self-force. Also, with these functions the regularized elastodynamic Green operator is amenable to elementary functions.

\subsection{Expression for a regularized dislocation loop}
The dislocation motion along direction $\bbm$ is constrained on the plane of unit normal $\bn$. The time-dependent dislocation position is $\bs(t)=s(t)\bm$. The unit tangent vector is $\bxi=\bn\times\bbm=-\be^3$. The unit vectors along the axes of the Cartesian coordinate system are $m_i=\delta_{i1}$, $n_i=\delta_{i2}$, and $\be^3_i=\delta_{i3}$.  The dislocation current is
\begin{align}
\label{eq:current}
I_{ij}(\br,t)&=b_i\varepsilon_{jkl}\oint_{\mathcal{L}(t)}\delta_\varepsilon^{(3)}(\br-\bs)V_k(\bs,t)\dd s_l,
\end{align}
where $\bV(\bs,t)$ is the velocity vector.\footnote{The convention for the indices is the transpose of, e.g., the one used by \citet{KRON58}, Kosevich and Mura, whereas Rogula defines the dislocation density with a negative sign.} If a curvilinear coordinate $\lambda$ is used to integrate along the loop, so that $\bs=\bs(\lambda,t)$, then $\bV(\bs,t)=\bV(\lambda,t)=\partial_t\bs(\lambda,t)$.

The density and current are related by the conservation equation for the Burgers vector \citep{KOSE62,KOSE79}
\begin{align}
\dot{\alpha}_{ij}(\br,t)+\epsilon_{jkl}I_{ik,l}(\br,t)=0.
\end{align}

The stress $\sigma_{ij}$ follows from the elastic distortion  $\beta_{ij}$ by Hooke's law $\sigma_{ij}(\br,t)=c_{ijkl}\beta_{kl}$, where the isotropic elasticity tensor reads in terms of the Lam\'e elastic moduli $\lambda$ and $\mu$,
\begin{align}
c_{ijkl}
&=\lambda\delta_{ij}\delta_{kl}+\mu(\delta_{ik}\delta_{jl}+\delta_{il}\delta_{jk}).
\end{align}
The elastic distortion and the material velocity are determined by \citep{MURA63,ROGU65}
\begin{subequations}
\begin{align}
\label{eq:dist0}
\beta_{ij}(\br,t)&=\int_{-\infty}^t\dd t'\int\dd^3r'
\left[\epsilon_{jln}\,G_{im,k}(\br-\br',t-t')\,c_{mkpl}\,\alpha_{pn}(\br',t')+\rho\,\dot{G}_{im}(\br-\br',t-t')I_{mj}(\br',t')\right],\\
\label{eq:v0}
v_i(\br,t)&=\int_{-\infty}^t\dd t'\int\dd^3r' G_{ij,k}(\br-\br',t-t')\,c_{jklm}I_{lm}(\br',t').
\end{align}
\end{subequations}
The expressions involve the elastodynamic Green operator $G_{ij}(\br,t)$, which is a generalized function \citep{PELL15}. Substituting \eqref{eq:loopdendyn} and \eqref{eq:current} into this equation and introducing the regularized 3D Green operator
\begin{align}
G^\varepsilon=G*\delta_\varepsilon^{(3)},
\end{align}
yields
\begin{subequations}
\begin{align}
\label{eq:distdyn0}
\beta_{ij}(\br,t)
&=b_p\epsilon_{jln}\int_{-\infty}^t\dd t'\oint_{\mathcal{L}(t')}
\left[G^\varepsilon_{im,k}(\br-\bs,t-t')c_{mkpl}+\rho\,\dot{G}^\varepsilon_{ip}\bigl(\br-\bs,t-t'\bigr)V_l(\bs,t')\right]\dd s_n\\
\label{eq:veldyn0}
v_i(\br,t)&=b_p\epsilon_{mln}\int_{-\infty}^t\dd t'\oint_{\calL(t')}G^\varepsilon_{ij,k}(\br-\bs,t-t')\,c_{jkmp}V_l(\bs,t')\,\dd s_n.
\end{align}
\end{subequations}
The 3D regularized Green operator is computed explicitly elsewhere. It is nowhere singular. The reader is referred to the above reference for its analytical expression in the two-dimensional problem examined in the next section.

The DPK force is given by the Peach-Koehler formula \citep{PEAC50} generalized to a spread dislocation density \citep{ROGU65}
\begin{align}
\label{eq:pkforce}
f^{\text{PK}}_i(t)
&=\epsilon_{ijk}\int \dd^3r\,\sigma_{lj}(\br,t)\alpha_{lk}(\br,t)=\epsilon_{ijk}c_{ljmn}\int \dd^3r\,\beta_{mn}(\br,t)\alpha_{lk}(\br,t).
\end{align}
Our line orientation convention is that of Fig.\ 2 in the PK reference; see, e.g., \cite{LUBA19} for a review.
Note that even in the presence of a `climb' Burgers vector component, the usual form of the elastic PK force has been employed. In writing so, we adhere to \citet{LOTH67} and take the stress-free crystal as the reference state \citep{DEWI68}, disregarding any osmotic force on vacancies; e.g., \citep{KEBO14}. Indeed in the present context a `climb' component simply stands for a step at a partial dislocation that bounds a stacking fault. If, on the contrary, one wishes to follow \citet{WEER65} (see also \citet[p.\ 118]{LAND86}), it is necessary to replace in \eqref{eq:pkforce} the stress by its deviator $\sigma'_{ij}=\sigma_{ij}-\delta_{ij}\sigma_{kk}/3$, and consequently to replace $c_{ljmn}$ by the correspondingly modified elastic tensor $c'_{ljmn}=c_{ljmn}-c_{kkmn}\delta_{lj}/3$. For either choice, the in-plane components of the force for the `screw' and 'edge' dislocations remain the same (see below).

Since the core-shape function obeys $\delta_\varepsilon^{(3)}*\delta_\varepsilon^{(3)}=\delta_{2\varepsilon}^{(3)}$, one gets
\begin{align}
\label{eq:fpk}
f^{\text{PK}}_i(t)&=b_l b_u\epsilon_{ijk}\epsilon_{nps}c_{ljmn}\int_{-\infty}^t\dd t'\oint_{\mathcal{L}(t)}\dd s_k\oint_{\mathcal{L}(t')}\dd s'_s
\left[G^{2\varepsilon}_{mr,o}(\bs-\bs',t-t')c_{roup}+\rho\,\dot{G}^{2\varepsilon}_{mu}\bigl(\bs-\bs',t-t'\bigr)V_p(\bs',t')\right].
\end{align}

\subsection{Straight regularized dislocation}
\label{sec:straightreg}
We now apply the above to a straight dislocation, for which one must take $\bs(t)=\xi(t)\bbm+z\be^3$, where $\xi(t)$ is the dislocation position along axis $Ox$. The line integration element becomes $\dd s_i=\delta_{i3}\dd z$, and $V_p(\bs,t)=\dot{\xi}(t)\delta_{1p}$. The loop integrations reduce to integrations over the $z$ coordinate. In the limit of an infinite dislocation, boundary contributions at $z=\pm\infty$ can be ignored. We go to 2D coordinates in the plane normal to the dislocation line. Introducing the two-dimensional regularized Green operator
\begin{align}
G^{(2)\varepsilon}_{ij}(x,y,t)&=\int_{-\infty}^{+\infty}\dd z G^{(2)\varepsilon}_{ij}(x\bbm+y\bn+z\be^3,t),
\end{align}
the force \emph{per unit dislocation line} is then
\begin{align}
\label{eq:selfforcebase}
f^{\text{PK}}_i(t)&=b_l b_u\epsilon_{ij3}\epsilon_{np3} c_{ljmn}\int_{-\infty}^t\dd t'
\left[G^{(2)2\varepsilon}_{mr,o}(\xi(t)-\xi(t'),0,t-t')c_{roup}+\rho\,\dot{G}^{(2)2\varepsilon}_{mu}\bigl(\xi(t)-\xi(t'),
0,t-t'\bigr)\dot{\xi}(t')\delta_{p1}\right].
\end{align}
The force component $f^{\text{PK}}_3(t)$ along the straight dislocation line vanishes, as the presence of the factor $\epsilon_{ij3}$ shows. The only  nonzero components of the regularized two-dimensional Green operator, which does not depend on the third coordinate, are $G_{11}$, $G_{12}=G_{21}$, $G_{22}$ and $G_{33}$. Computing the index contractions, and omitting for brevity all the obvious superscripts and arguments, one finds that:
\begin{itemize}
\item `Screw': $\bb=[0,0,-b]$.
\begin{align}
\label{eq:screw12}
f^{\rm s}_1&=-b^2\mu^2\int_{-\infty}^t \dd t'\left[G_{33,1} + \frac{1}{\cS^2}\dot{G}_{33}\dot{\xi}(t')\right],\qquad
f^{\rm s}_2=-b^2\mu^2\int_{-\infty}^t \dd t'\,G_{33,2};
\end{align}
\item `Glide' edge: $\bb=[b,0,0]$.
\begin{subequations}
\begin{align}
\label{eq:ge1}
f^{\rm g}_1&=-b^2\mu^2\int_{-\infty}^t \dd t'\left[
\phi^2G_{11,1}
+\left(\phi^2-3\right)G_{12,2}
-G_{22,1}
+\frac{1}{\cS^2}\dot{G}_{11}\dot{\xi}(t')\right],\\
\label{eq:ge2}
f^{\rm g}_2&=-b^2\mu^2\int_{-\infty}^t \dd t'\left[
\phi^2G_{11,2}
+\phi^2\left(3-\phi^2\right)G_{12,1}
-\left(\phi^2-2\right)^2G_{22,2}+\frac{1}{\cS^2}\left(2-\phi^2\right)\dot{G}_{12}\dot{\xi}(t')\right];
\end{align}
\end{subequations}
\item `Climb' edge: $\bb=[0,b,0]$.
\begin{subequations}
\begin{align}
\label{eq:ce1}
f^{\rm c}_1&=b^2\mu^2\int_{-\infty}^t \dd t'\left[
 \left(\phi^2-2\right)^2 G_{11,1}+\phi^2\left(\phi^2-3\right)G_{12,2}-\phi^2 G_{22,1}
-\frac{\phi^2}{\cS^2}\dot{G}_{22}\dot{\xi}(t')\right],\\
\label{eq:ce2}
f^{\rm c}_2&=b^2\mu^2\int_{-\infty}^t \dd t'\left[
G_{11,2}+\left(3-\phi^2\right)G_{12,1}-\phi^2 G_{22,2}+\frac{1}{\cS^2}\dot{G}_{12}\dot{\xi}(t')\right].
\end{align}
\end{subequations}
\end{itemize}
Because if $y=0$, $G_{33,2}=G_{11,2}=G_{12,1}=G_{22,2}=\dot{G}_{12}=0$, the above reduces to
\begin{subequations}
\begin{itemize}
\item `Screw':
\begin{align}
\label{eq:f1s}
f^{\rm s}_1&=-b^2\mu^2\int_{-\infty}^t \dd t'\left[G_{33,1}+\frac{1}{\cS^2}\dot{G}_{33}\dot{\xi}(t')\right],\qquad f^{\rm s}_2=0;
\end{align}
\item `Glide' edge:
\begin{align}
\label{eq:f1g}
f^{\rm g}_1&=-b^2\mu^2\int_{-\infty}^t \dd t'\left[\phi^2 G_{11,1}+\left(\phi^2-3\right)G_{12,2}-G_{22,1}+
   \frac{1}{\cS^2}\dot{G}_{11}\dot{\xi}(t')\right],\qquad f^{\rm g}_2=0;
\end{align}
\item `Climb' edge:
\begin{align}
\label{eq:f1c}
f^{\rm c}_1&=b^2 \mu^2\int_{-\infty}^t \dd t'\left[\left(\phi^2-2\right)^2G_{11,1}
+\phi^2\left(\phi^2-3\right)G_{12,2}-\phi^2 G_{22,1}
-\frac{\phi^2}{\cS^2}\dot{G}_{22}\dot{\xi}(t')\right],\qquad
f^{\rm c}_2=0.
\end{align}
\end{itemize}
\end{subequations}
with $f_2^{\rm s,g,c}=0$. Thus, the self-force only acts in the direction of motion. This result is independent of the regularization procedure. Had the \citet{WEER65} modification of the self-force been used (with \eqref{eq:fpk} written with $c'_{ljmn}$ rather than with $c_{ljmn}$), then Eqs.\ \eqref{eq:ge2} and \eqref{eq:ce1} would have been modified, inducing the following changes in equations \eqref{eq:f1g}--\eqref{eq:f1c}: $f^{\rm g}_2=(2/3)f^{\rm g}_1$; $f^{\rm c}_1=-(2/3)b^2 \mu^2\int \dd t'\,[(\phi^2-2)G_{11,1}+(\phi^2+2)G_{12,2}+2G_{22,1}+(2/\cS^2)\dot{G}_{22}\dot{\xi}(t')]$. Thus, insofar as only the in-plane force component for the screw and the glide edge dislocations is of interest, this modification is of no consequence. However, it affects the normal component of the force for the glide-edge dislocation, and both components of the force for the climb-edge dislocation. Weertman's modification is not further considered hereafter.

\section{Self-force in `mass' and `impulsion' forms}
\label{sec:massimp}
In this section, the self-force is computed explicitly for arbitrary motion under specially appealing forms.
\subsection{`Mass' form of the self-force with partial mass functions}
With the regularization defined by \eqref{eq:delta2}, the 2D regularized elastodynamic Green operator $G^{(2)\varepsilon}=G^{(2)}\star\delta^{(2)}_\varepsilon$ has been computed by \citet{PELL15}. With $\br=(x,y)$, its nonzero components read
\begin{subequations}
\begin{align}
\label{eq:green2D}
G^{(2)\varepsilon}_{ij}(\br,t)&=\frac{\theta(t)}
{4\pi\rho}\Re\left\{
\sum_{{\rm P}={\rm S,L}}
\frac{1}{\cP\sqrt{\cP^2 t^2-r^2}}\left[\delta_{ij}\pm \frac{1}{r^2}\left(2\cP^2t^2-r^2\right)
\left(\delta_{ij}-2\frac{r_i r_j}{r^2}\right)\right]\right\}_{\cP t\to\cP t+\ii\varepsilon} i,j=1,2,\\
G^{(2)\varepsilon}_{33}(\br,t)&=\frac{\theta(t)}
{2\pi\rho}\Re\frac{1}{\cS\sqrt{(\cS t+\ii\varepsilon)^2-r^2}},
\end{align}
\end{subequations}
where $\theta(t)$ is the Heaviside function, and where the signs $\pm$ apply to the S and L wave-terms, respectively. These expressions indicate that in two dimensions computing the above convolution essentially amounts to replacing in the standard Green operator $G^{(2)}$ \citep[p.\ 412]{ERIN75} the propagation distances $\cS t$ and $\cL t$ by $\cS t+\ii \varepsilon$ and $\cL t+\ii \varepsilon$, respectively, and taking the real part. By homogeneity, the standard Green operator depends on time only via the groups $r/(\cS t)$ and $r/(\cL t)$. Consequently, expressions such as $[\xi(t)-\xi(t')]/[\cSL(t-t')]$ within the standard operator are transformed by regularization into $[\xi(t)-\xi(t')]/[\cSL(t-t')+\ii\varepsilon]$ in the regularized one \eqref{eq:selfforcebase}. This makes it desirable to introduce the following complex-valued `mean velocities' between instants $t$ and $t'$:
\begin{align}
\label{eq:vbar}
\overline{v}_{\rm S,L}(t,t')&=\frac{\cSL\ \Delta\xi(t,t')}{\cSL(t-t')+\ii a},
\end{align}
where
\begin{align}
\Delta\xi(t,t')&=\xi(t)-\xi(t'),
\end{align}
and account has been made of the doubling of the regularizing length as $a=2\varepsilon$ -- the core diameter -- in the Green operator within \eqref{eq:selfforcebase}. The elements $G^{(2)2\varepsilon}_{ij,k}(\xi(t)-\xi(t'),0,t-t')$ of interest, expressed in terms of $\overline{v}_{\rm S,L}$, are listed in Appendix \ref{sec:elements}. By contrast with the flat-core limits \eqref{eq:selfforceflat}, one now has
\begin{align}
\label{eq:limitsiso}
\lim_{t'\to t^-}\vSL(t,t')&=0,\qquad
\lim_{t'\to -\infty}\vSL(t,t')=\dot{\xi}(-\infty)+\ii 0^+.
\end{align}
Considering that: (i) the regularized Green operator can be written as a sum of separate operators relative to each partial wave $\text{P}=\text{S}, \text{L}$; (ii) each operator can be written as a function of the variable $\vSL(t,t')$, times a factor $1/\Delta\xi^2(t,t')$, there exists tensor functions $H_{ij}^{\rm S,L}(z)$ and $H_{ijk}^{\rm S,L}(z)$ of a single complex variable $z$ (hereafter distinguished only by their number of indices) such that
\begin{subequations}
\begin{align}
\label{eq:dtHdef}
\dot{G}_{ij}^{(2)2\varepsilon}(\xi(t)-\xi(t'),0,t-t')&=\frac{\cS^2}{2\pi\mu}\frac{1}{\Delta\xi^2(t,t')}\Re\sum_{\rm P=S,L}H^{\rm P}_{ij}\left(\frac{\vP(t,t')}{\cP}\right),\\
\label{eq:gradHdef}
G_{ij,k}^{(2)2\varepsilon}(\xi(t)-\xi(t'),0,t-t')&=
\frac{\cS}{2\pi\mu}\frac{1}{\Delta\xi^2(t,t')}\Re\sum_{\rm P=S,L}H^{\rm P}_{ijk}\left(\frac{\vP(t,t')}{\cP}\right),
\end{align}
\end{subequations}
The $H^{\rm L,S}(z)$ tensor elements of interest are read directly from Eqs.\ \eqref{eq:dtg11}--\eqref{eq:g331} of Appendix \ref{sec:elements}. It can be verified that the following equal-time limits hold:
\begin{subequations}
\label{eq:eqtime}
\begin{align}
\lim_{t'\to t}\dot{G}_{ij}^{(2)2\varepsilon}(\xi(t)-\xi(t'),0,t-t')&=\frac{\cS^2}{2\pi\mu a^2}\delta_{ij},\\
\lim_{t'\to t}\dot{G}_{ij,k}^{(2)2\varepsilon}(\xi(t)-\xi(t'),0,t-t')&=0.
\end{align}
\end{subequations}
These well-behaved limits stem from a cancellation of inessential singularities between partial-wave terms. However, the derivation to follow requires equal-time limits relative to \emph{each wave} to be well-defined. To this aim, some functions $H^{\rm P}(z)$ are modified by subtracting out their leading-order term $z^\alpha$ with $\alpha\leq 0$, when $z\to 0$, which leaves unchanged the overall expressions \eqref{eq:dtHdef}--\eqref{eq:gradHdef}. These modified $H$-functions are listed in Appendix \ref{sec:elements}.

Thus, using the derivative $(\dd/\dd t')1/\Delta\xi(t,t')=1/\Delta\xi^2(t,t')$, and the identities
\begin{align}
\frac{1}{\Delta\xi(t,t')}&=\frac{1}{\cP(t-t')+\ii\,a}\left(\frac{\cP}{\vP}\right),\qquad
\frac{1}{\Delta\xi^2(t,t')}=\frac{1}{\cP}\left(\frac{\cP}{\vP}\right)^2\frac{\dd}{\dd t'}\frac{1}{\cP(t-t')+\ii\,a},
\end{align}
one gets,
\begin{align}
&G^{(2)2\varepsilon}_{mr,o}(\xi(t)-\xi(t'),0,t-t')c_{roup}+\rho\,\dot{G}^{(2)2\varepsilon}_{mu}\bigl(\xi(t)-\xi(t'),
0,t-t'\bigr)\delta_{p1}\dot{\xi}(t')\nonumber\\
&=\frac{1}{2\pi\mu}\Re\sum_{\rm P=S,L}\frac{1}{\Delta\xi^2(t,t')}
\left[\cS\,
H^{\rm P}_{mro}\left(\frac{\vP}{\cP}\right)c_{roup}
+\rho\,\cS^2 H^{\rm P}_{mu}\left(\frac{\vP}{\cP}\right)\delta_{p1}\dot{\xi}(t')\right]\nonumber\\
&=\frac{1}{2\pi\mu}\Re\sum_{\rm P=S,L}
\left[\left(\frac{\dd}{\dd t'}\frac{1}{\cP(t-t')+\ii\,a}\right)\frac{\cS}{\cP}\frac{\cP^2}{\vP^2}
H^{\rm P}_{mro}\left(\frac{\vP}{\cP}\right)c_{roup}
+\mu\,H^{\rm P}_{mu}\left(\frac{\vP}{\cP}\right)\delta_{p1}\frac{\dd}{\dd t'}\frac{1}{\Delta\xi(t,t')}\right]\nonumber\\
&=\frac{\dd B_{mup}}{\dd t'}(t,t')-\frac{1}{2\pi\mu}\Re\sum_{\rm P=S,L}
\frac{\cS}{\cP}\left\{\frac{1}{\cP(t-t')+\ii\,a}\frac{\dd}{\dd t'}\left[\frac{\cP^2}{\vP^2}
H^{\rm P}_{mro}\left(\frac{\vP}{\cP}\right)\right]c_{roup}
+\frac{\mu}{\Delta\xi}\left[\frac{\dd}{\dd t'}H^{\rm P}_{mu}\left(\frac{\vP}{\cP}\right)\right]\delta_{p1}\right\},\nonumber\\
&={}\frac{\dd B_{mup}}{\dd t'}(t,t')\nonumber\\
\label{eq:allg}
&{}-\frac{1}{2\pi\mu}\Re\sum_{\rm P=S,L}
\frac{1}{\cP(t-t')+\ii\,a}\frac{1}{\cP}\left\{\frac{\cS}{\cP}\left[z^{-2}H^{\rm P}_{mro}(z)\right]'c_{roup}
+\mu\,z^{-1}\left[H^{\rm P}_{mu}(z)\right]'\delta_{p1}\right\}_{z=\frac{\vP(t,t')}{\cP}}\frac{\dd \vP}{\dd t'}(t,t').
\end{align}
where the prime stands for a derivative with respect to $z$, and
\begin{align}
B_{mup}(t,t')&=\frac{1}{2\pi\mu}\frac{1}{\Delta\xi(t,t')}\Re\sum_{\rm P=S,L}
\left[\frac{\cS}{\vP}H^{\rm P}_{mro}\left(\frac{\vP}{\cP}\right)c_{roup}+\mu\,H^{\rm P}_{mu}\left(\frac{\vP}{\cP}\right)\delta_{p1}\right].
\end{align}
Thanks to the use of the modified $H$-functions, the $B_{mup}(t,t')$ terms can be shown to vanish at the boundaries $t'\to-\infty$ and $t'=t$ of the time integral, so that the leftmost time-derivative in \eqref{eq:allg} plays no part. This step is key to moving forward.

It follows that upon introducing \emph{partial mass functions relative to each wave} defined as
\begin{align}
\label{eq:partm}
\widetilde{m}_{\rm P}(z)&=\frac{b_l b_u}{4\pi\mu}\epsilon_{np3}c_{l2mn}\frac{1}{\cP^2}\left\{\frac{\cS}{\cP}\left[z^{-2} H^{\rm P}_{mro}(z)\right]'c_{roup}+\mu\,z^{-1}\left[H^{\rm P}_{mu}(z)\right]'\delta_{p1}\right\},
\end{align}
the substitution of expression \eqref{eq:allg} into \eqref{eq:selfforcebase} eventually yields the in-plane component of the self-force in `mass form' as
\begin{align}
\label{eq:massform}
f^{\text{PK}}_1(t)&=-2\Re\int_{-\infty}^t \dd t'\,\sum_{\rm P=S,L}
\frac{\cP}{\cP(t-t')+\ii\,a}\widetilde{m}_{\rm P}\left(\frac{\vP(t,t')}{\cP}\right)\frac{\dd \vP}{\dd t'}(t,t').
\end{align}

If we formally let $a=0$, both mean velocities $\vSL$ reduce to $\overline{v}$, Eq.\ \eqref{eq:meanveltemplate}, and the force reduces to \eqref{eq:template} with the quantity
\begin{align}
\label{eq:masssum}
m(v)=\sum_{\rm P=S,L}\widetilde{m}_{\rm P}(v/\cP)
\end{align}
coinciding with the usual (prelogarithmic) mass function of a Volterra dislocation. The partial mass functions \eqref{eq:partm}, evaluated for the three dislocation characters with the help of \eqref{eq:H11}--\eqref{eq:H331}, read
\begin{itemize}
\item \emph{Screw dislocation:}
\begin{subequations}
\begin{align}
\widetilde{m}_{\rm S}^{\rm s}(z)&=\frac{w_0}{\cS^2}\left(1-z^2\right)^{-3/2},\qquad m^{\rm s}_{\rm L}(z)=0;
\end{align}
\end{subequations}
\item \emph{Glide-edge dislocation:}
\begin{subequations}
\begin{align}
\widetilde{m}_{\rm S}^{\rm g}(z)&=\frac{w_0}{\cS^2}
z^{-4}\left[\left(24-60\,z^2+46\,z^4-7\,z^6\right)\left(1-z^2\right)^{-5/2}-24\right],\\
\widetilde{m}_{\rm L}^{\rm g}(z)&=-4\frac{w_0}{\cS^2}
(\phi\,z)^{-4}\left[\left(6-9\,z^2+2\,z^4\right)\left(1-z^2\right)^{-3/2}-6\right];
\end{align}
\end{subequations}
\item \emph{Climb-edge dislocation:}
\begin{subequations}
\begin{align}
\widetilde{m}^{\rm c}_{\rm S}(z)&=-4\frac{w_0}{\cS^2}
z^{-4}\left[\left(6-9\,z^2+2\,z^4\right)\left(1-z^2\right)^{-3/2}-6\right],\\
\widetilde{m}^{\rm c}_{\rm L}(z)&=\frac{w_0}{\cS^2}(\phi\,z)^{-4}\left\{
\left[24-60\,z^2+2\left(24-2\phi^2+\phi^4\right)\,z^4+\phi^2\left(\phi^2-8\right)z^6\right]\left(1-z^2\right)^{-5/2}
-24\right\}.
\end{align}
\end{subequations}
\end{itemize}
These partial mass functions go to finite values as $z\to 0$. Combining the $\text{S}$ and $\text{L}$ functions according to \eqref{eq:masssum}, the mass expressions \eqref{eq:ms}--\eqref{eq:mc} of a Volterra dislocation are retrieved in each case.

\subsection{`Impulsion' form of the self-force with partial impulsion functions}
An 'impulsion'-form expression of the self-force \citep{PELL14} is now derived using partial impulsion functions $\widetilde{p}^{\rm P}(z)$ defined as
\begin{align}
\widetilde{p}_{\rm P}(z)&=\cP\int_0^z\dd z\,\widetilde{m}_{\rm P}(z).
\end{align}
Since $p'(v)=m(v)$, with $m(v)$ given by \eqref{eq:masssum}, the usual impulsion function is retrieved as
\begin{align}
\label{eq:psum}
p(v)&=\sum_{\rm P=S,L}\widetilde{p}_{\rm P}(v/\cP).
\end{align}
Consequently,
\begin{align}
\frac{\dd}{\dd t'}\widetilde{p}_{\rm P}(\vP(t,t')/\cP)&=\widetilde{m}_{\rm P}\left(\vP(t,t')/\cP\right)\frac{\dd\vP}{\dd t'}(t,t').
\end{align}
Using this derivative, Eq.\ \eqref{eq:massform} is integrated by parts, with vanishing boundary contributions. The following \emph{`ìmpulsion' form of the self-force} obtains:
\begin{align}
\label{eq:impulsionform}
f^{\text{PK}}_1(t)&=2\Re\int_{-\infty}^t \dd t'\,\sum_{\rm P=S,L}
\frac{\cP^2}{[\cP(t-t')+\ii\,a]^2}\widetilde{p}_{\rm P}\left(\frac{\vP(t,t')}{\cP}\right).
\end{align}
The partial impulsion functions for the three characters read:
\begin{itemize}
\item \emph{Screw dislocation:}
\begin{subequations}
\begin{align}
\widetilde{p}_{\rm S}^{\rm s}(z)&=\frac{w_0}{\cS}z\left(1-z^2\right)^{-1/2},\qquad \widetilde{p}^{\rm s}_{\rm L}(z)=0;
\end{align}
\end{subequations}
\item \emph{Glide-edge dislocation:}
\begin{subequations}
\begin{align}
\widetilde{p}_{\rm S}^{\rm g}(z)&=-\frac{w_0}{\cS}z^{-3}
\left[\left(1-z^2\right)^{-3/2}\left(8-12\,z^2+2\,z^4+z^6\right)-8\right],\\
\widetilde{p}_{\rm L}^{\rm g}(z)&=4\frac{w_0}{\cS}(\phi\,z)^{-3}
\left[\left(1-z^2\right)^{-1/2}\left(2-z^2\right)-2\right];
\end{align}
\end{subequations}
\item \emph{Climb-edge dislocation:}
\begin{subequations}
\begin{align}
\widetilde{p}^{\rm c}_{\rm S}(z)&=4\frac{w_0}{\cS}z^{-3}
\left[\left(1-z^2\right)^{-1/2}\left(2-z^2\right)-2\right],\\
\widetilde{p}^{\rm c}_{\rm L}(z)&=-\frac{w_0}{\cS}(\phi\,z)^{-3}
\left\{\left(1-z^2\right)^{-3/2}
\left[8-12\,z^2+2\phi^2\left(2-\phi^2\right)\,z^4+\phi^4 z^6\right]-8\right\}.
\end{align}
\end{subequations}
\end{itemize}
Substituting these functions into \eqref{eq:psum} yields Eqs.\ \eqref{eq:ps}-\eqref{eq:pc}.

\subsection{Partial energy-related functions}
The next Section requires for $\text{P}=\text{S},\text{L}$, partial Lagrangian functions $\wL_{\text{P}}$ defined such that
\begin{align}
\label{eq:partlag}
\wL_{\text{P}}'(z)&=\cP\,\widetilde{p}_{\text{P}}(z),\quad\text{and}\quad L(v)=\wL_{\text{S}}(v/\cS)+\wL_{\text{L}}(v/\cL),
\end{align}
namely,
\begin{itemize}
\item \emph{Screw dislocation:}
\begin{subequations}
\begin{align}
\wL^{\rm s}_{\rm S}(z)&=-w_0\,\left(1-z^2\right)^{1/2},\qquad \wL^{\rm s}_{\rm L}(z)=0;
\end{align}
\end{subequations}
\item \emph{Glide-edge dislocation:}
\begin{subequations}
\begin{align}
\wL^{\rm g}_{\rm S}(z)&=4 w_0\,\red{z^{-2}}\left[\red{\left(1-z^2/2\right)^{2}}\left(1-z^2\right)^{-1/2}-1\right],\\
\wL^{\rm g}_{\rm L}(z)&=-4 w_0\,\red{(\phi\,z)^{-2}}\left[\left(1-z^2\right)^{1/2}-1\right];
\end{align}
\end{subequations}
\item \emph{Climb-edge dislocation:}
\begin{subequations}
\begin{align}
\wL^{\rm c}_{\rm S}(z)&=-4 w_0\,\red{z^{-2}}\left[\left(1-z^2\right)^{1/2}-1\right],\\
\wL^{\rm c}_{\rm L}(z)&=4 w_0\,\red{(\phi\,z)^{-2}}\left[\left(1-\phi^2 z^2/2\right)^2\left(1-z^2\right)^{-1/2}-1\right].
\end{align}
\end{subequations}
\end{itemize}
Likewise, partial energy functions are introduced as
\begin{align}
\widetilde{W}_{\text{P}}(z)&=\cP\,z\,\widetilde{p}_{\text{P}}(z)-\widetilde{L}_{\text{P}}(z).
\end{align}
From these, partial dislocation kinetic energies are defined as
\begin{align}
\label{eq:parkin}
\widetilde{K}_{\text{P}}(z)&=\widetilde{W}_{\text{P}}(z)-\widetilde{W}_{\text{P}}(0).
\end{align}

\section{Initial conditions and particular motions}
\label{sec:initcond}
In this section, the self-force is computed analytically for particular motions from the above integral representations, by adapting the method employed in \citep{PELL12,PELL14}.

\subsection{Steady motion on a time interval}
If at time $t$ the dislocation has undergone steady motion at speed $v$ (possibly, $v=0$) during the past time interval $t'\in[t_1,t_2]$, where $-\infty\leq t_1< t_2\leq t$, the contribution $\smash{f_{1\,[t_1,t_2]}^{\text{PK}}(t)}$ of this interval to the self-force \eqref{eq:massform} is as follows. The time variable $t'$ is traded for $u=\overline{v}_{\rm P}$ as the integration variable, letting $u_{{\rm P}\,1,2}(t)=\overline{v}_{\rm P}(t,t_{1,2})$. Then,
\begin{align}
f_{1\,[t_1,t_2]}^{\text{PK}}(t)
&=-2\Re\int_{t_1}^{t_2}\sum_{\rm P=S,L}\frac{\cP\,\dd t'}{\cP(t-t')+\ii\,a}\widetilde{m}_{\text{P}}(\vP/\cP)\frac{\dd\vP}{\dd t'}(t,t)\nonumber\\
&=-2\Re\sum_{\rm P=S,L}\frac{\cP}{v(\cP\,t+\ii\,a)-\cP\,\xi(t)}\int_{u_{\text{P}\,1}(t)}^{u_{\text{P}\,2}(t)}\dd u\,(v-u)\widetilde{m}_{\text{P}}(u/\cP).
\end{align}
The indefinite integral (by one partial integration)
\begin{align}
\int\dd u\,(v-u)\widetilde{m}_{\rm P}(u/\cP)&=(v-u)\widetilde{p}_{\text{P}}(u/\cP)+\widetilde{L}_{\text{P}}(u/\cP)=v\, \widetilde{p}_{\text{P}}(u/\cP)-\widetilde{W}_{\text{P}}(u/\cP),
\end{align}
gives
\begin{align}
\label{eq:steadymotion}
f_{1\,[t_1,t_2]}^{\text{PK}}(t)&=-2\Re\sum_{\rm P=S,L}\frac{\cP}{v(\cP t+\ii\,a)-\cP\xi(t)}
\left\{v\left[\widetilde{p}_{\text{P}}\left(u_{\text{P}\,2}/\cP\right)-\widetilde{p}_{\text{P}}\left(u_{\text{P}\,1}/\cP\right)\right]
-\left[\widetilde{W}_{\text{P}}\left(u_{\text{P}\,2}/\cP\right)
-\widetilde{W}_{\text{P}}\left(u_{\text{P}\,1}/\cP\right)\right]\right\}.
\end{align}

\begin{figure}
\begin{center}
\includegraphics[width=16cm]{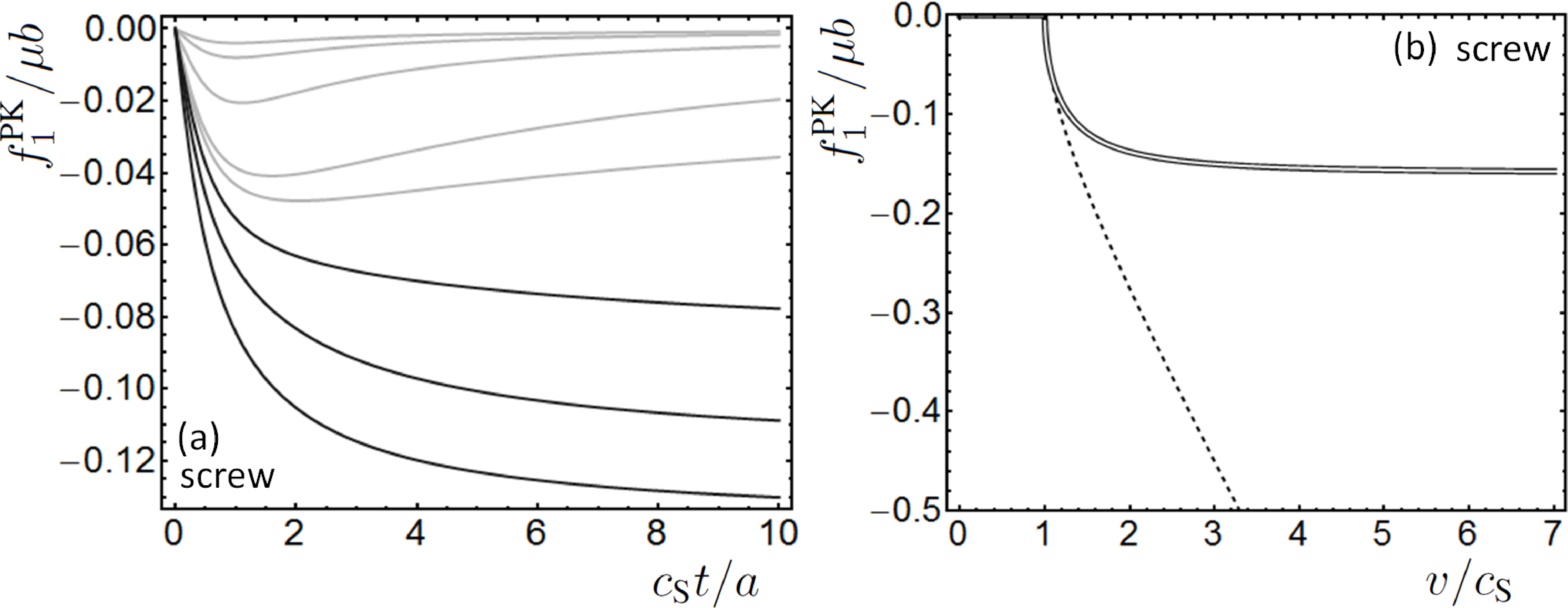}\\
\includegraphics[width=16cm]{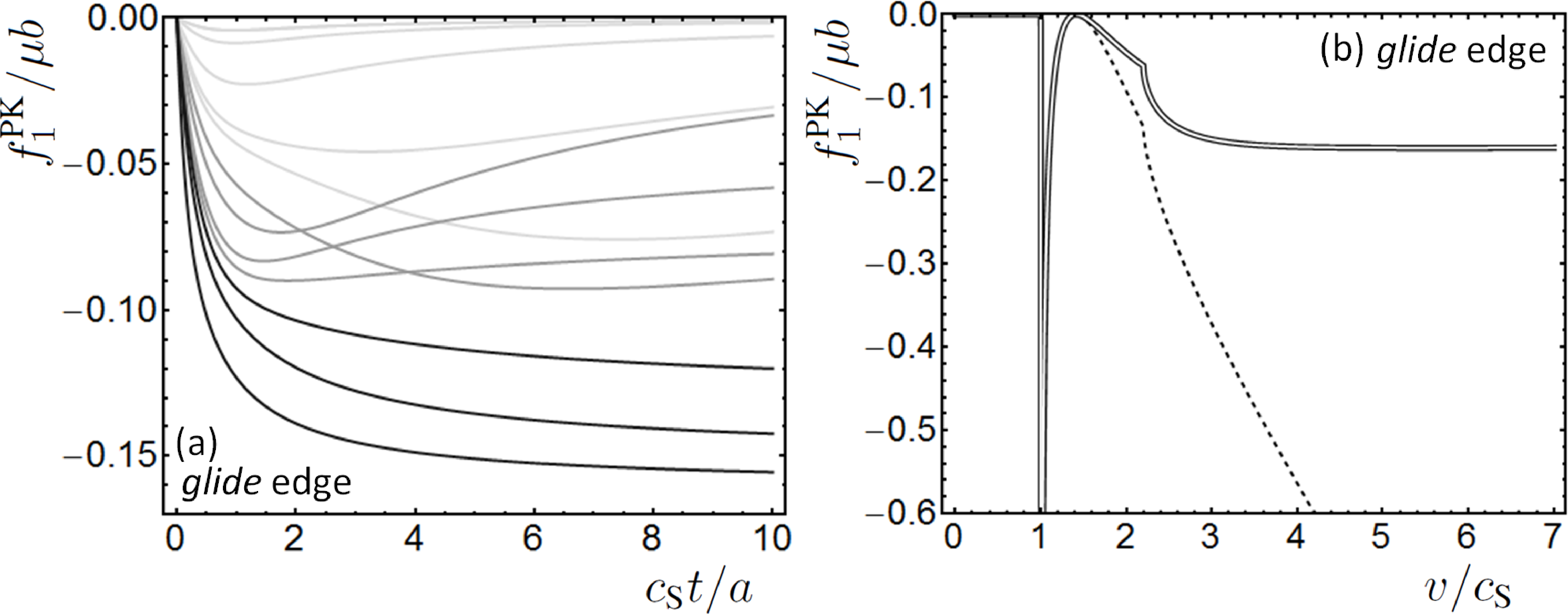}\\
\includegraphics[width=16cm]{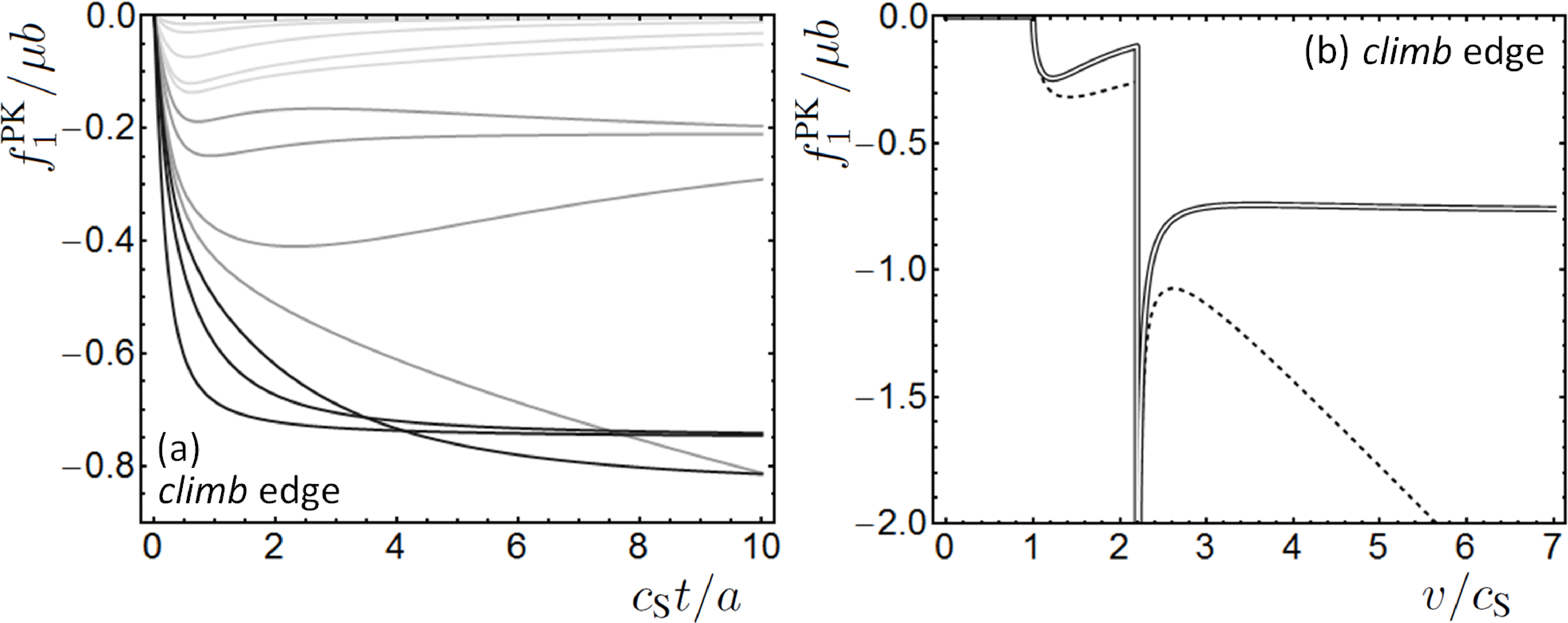}
\end{center}
\caption{\label{fig:fig1} Dynamic Peach-Koehler self-force for screw (top), glide-edge (middle), and climb-edge (bottom) dislocations. Subplots
(a): dynamics as a function of time, Eq.\ \eqref{eq:jumpfromrest}, for jump from rest to a steady-state of speed $v$, for $v/\cS=0.1$, $0.2$, $0.5$, $0.9$, $1.0$, $1.2$, $1.5$, $2.0$ (screw; light grey for $v/\cS\leq 1$ and black otherwise), and \red{$v/\cS=0.1$, $0.2$, $0.5$, $0.8$, $0.9$, $1.2$, $1.5$, $2.0$, $2.2$, $2.5$, $3.0$, $5.0$} (glide and climb edges; light grey for $v\cS\leq 1$, medium grey for $1<v/\cS\leq\cL/\cS$, and black otherwise). Subplots (b): asymptotic force \eqref{eq:jumpfromrestlt} (thick, black) with overlaid Eq.\ (3.28) from \citep{PELL18} (thin, white), and result for a flat-core dislocation (dashed; see text). Consideration of a fast `climb-edge' dislocation is speculative (see text).}
\end{figure}

\subsection{Dislocation starting from rest}
Now, if theF dislocation is at rest for $t<0$, and dynamic motion begins at $t=0$, the effect on the self-force of the dynamic relaxation (`erasure') of the initial static field of the dislocation at rest \citep{PELL14,LAZA16} is expressed by the contribution of $f_{1\,[-\infty,0]}^{\text{PK}}(t)$, computed from \eqref{eq:steadymotion} with $v=0$, and $u_{\text{P}\,1}(t)=0$ and $u_{\text{P}\,2}(t)=\cP\xi(t)/(\cP t+\ii a)$. Then,
\begin{align}
f_{1\,[-\infty,0]}^{\text{PK}}(t)
&=-\frac{2}{\xi(t)}\Re\sum_{\rm P=S,L}\widetilde{K}_{\text{P}}\left(\frac{\xi(t)}{\cP\,t+\ii\,a}\right),
\end{align}
where $\widetilde{K}_{\text{P}}$ is the partial kinetic-energy function \eqref{eq:parkin}. It follows that quite generally for any motion starting from rest
\begin{align}
\label{eq:motionfromrest}
f_1^{\text{PK}}(t)
&=-\frac{2}{\xi(t)}\Re\sum_{\rm P=S,L}\widetilde{K}_{\text{P}}\left(\frac{\xi(t)}{\cP\,t+\ii\,a}\right)-2\Re\int_{0}^{t}\sum_{\rm P=S,L}\frac{\cP\,\dd t'}{\cP(t-t')+\ii\,a}\widetilde{m}_{\text{P}}\left(\vP/\cP\right)\frac{\dd\vP}{\dd t'}(t,t').
\end{align}

If, furthermore, the dislocation moves for $t>0$ at constant speed $v$ with $\xi(t)=v t$, we fall into the prototypal case of a dislocation jumping instantaneously from rest to steady motion. Then, combining \eqref{eq:motionfromrest} with again \eqref{eq:steadymotion} used with $t_1=0$ and $t_2=t$, and noting that partial impulsions and kinetic energies vanish at the origin, one arrives at
\begin{align}
\label{eq:jumpfromrest}
f_1^{\text{PK}}(t)
&=\frac{2}{a}\Im\sum_{\rm P=S,L}\left[\cP\,\widetilde{p}_{\text{P}}(z)-z^{-1}\widetilde{K}_{\text{P}}(z)\right]_{z=v t/(\cP t+\ii\,a)}
=\red{\frac{2}{a}\Im\sum_{\rm P=S,L}\left[\frac{1}{z}\left(\widetilde{L}_{\text{P}}(z)+\widetilde{W}_{\text{P}}(0)\right)\right]_{z=v t/(\cP t+\ii\,a)}.}
\end{align}
Limits $a\to 0$ and $t\to+\infty$ do not commute. In the Volterra limit $a\to 0$ and subsonic regime $|v|<\cS$ (whence $\widetilde{K}_{\text{P}}$ and $\widetilde{p}_{\text{P}}$ are real) expression \eqref{eq:jumpfromrest} reduces (up to a change of sign; see present footnote 1)) to the self-force of the Volterra dislocation written in terms of the dislocation kinetic energy function \citep[Eq.\ (67)]{PELL12}; see also \citep{CLIF81}. In the long-time limit, expression \eqref{eq:jumpfromrest} reduces to the steady-state expression
\begin{align}
f_1^{\text{PK}}(v)
&=\frac{2}{a}\Im\sum_{\rm P=S,L}\frac{\cP}{v}\widetilde{L}_{\text{P}}\left(\frac{v}{\cP}-\ii\,0^\pm\right)
\label{eq:jumpfromrestlt}
=-\frac{2}{a}\Im\sum_{\rm P=S,L}\frac{\cP}{v}\widetilde{L}_{\text{P}}\left(\frac{v+\ii\,0^\pm}{\cP}\right)
\end{align}
where the sign of the infinitesimal imaginary part is that of $v$.

This asymptotic expression is different from that for a flat-core dislocation, namely (again up to an overall change of sign) \citep{PELL14},
\begin{align}
\label{eq:flatcore}
f_1^{\text{PK},\text{flat-core}}(v)=-\frac{2}{a}\Im L(v+\ii 0^+),
\end{align}
which involves the `standard' Lagrangian function \eqref{eq:partlag}${}_2$; see also Eqs.\ \eqref{eq:lags}--\eqref{eq:lagc}. The present regularization replaces this Lagrangian $L(v)$ by the composite function $(\cS/v)\widetilde{L}_{\text{S}}(v/\cS)+(\cL/v)\widetilde{L}_{\text{L}}(v/\cL)$ in \eqref{eq:jumpfromrestlt}. Thus, the regularization procedure deeply influences the self-force, and the `standard' Lagrangian function may not necessarily be always relevant. Irrespective of the model, the steady-state force is nonzero only for $|v|>\cS$, where it accounts for Cerenkov-induced radiative drag; e.g., \citep{PELL18}.

In Eq.\ (3.28) of the latter work the steady-state DPK force of the regularized dislocation was obtained by another method based on the Stroh formalism \citep{ANDE17}, for a medium of arbitrary elastic anisotropy, and for a more general \emph{elliptical} core of half semi-axes widths $a_\parallel$ and $a_\perp$. The regularization employed there reduces to the present one if $a_\parallel=a_\perp=a/2$. If Stroh vectors suitable to the isotropic medium are employed (not provided here for brevity), the $\mathop{\rm Arcsinh}(p_\alpha)$ term within the function $F^{(1)}(p_\alpha)$ in the said Eq.\ (3.28) turns out not to contribute ($p_\alpha$ denoting a Stroh eigenvalue). Expression \eqref{eq:jumpfromrestlt} then provides an equivalent, but notably simpler, form of the self-force. Thus, the present approach is fully consistent with earlier results, up to regularization-specific features.

Figure \ref{fig:fig1} illustrates these results for the three dislocation characters, and parameter values $a/b=1$ and $\cL/\cS=2.2$. Subplots (a) display the dynamics of the self-force for various terminal speeds, as a function of time, for a regularized dislocation jumping from rest to steady speed $v$. The self-force goes asymptotically to zero if $v$ is subsonic, and to a finite value when $v>\cS$. In the subsonic regime, the transient is a consequence of elastodynamic effective inertia. Asymptotic values are displayed versus $v$ in subplots (b), where they are compared with the result of \citep{PELL18} --with perfect coincidence, and with the force \eqref{eq:flatcore} for a flat-core dislocation. Although both models predict the same intersonic `radiation-free velocity' $v^*=\sqrt{2}\cS$ for the glide-edge dislocation \citep{ESHE49,GUMB99,GAOH99,MARK01}, there are striking differences: for $v>\cL$, the radiative drag of the flat-core dislocation increases with $v$, whereas that of the regularized dislocation saturates -- a behavior at odds with results from molecular dynamics. Over the whole velocity range, and irrespective of the character, the flat-core dislocation undergoes more radiative drag than the regularized dislocation, with more realistic features if $a$ is fixed.

\begin{figure}
\begin{center}
\includegraphics[width=16cm]{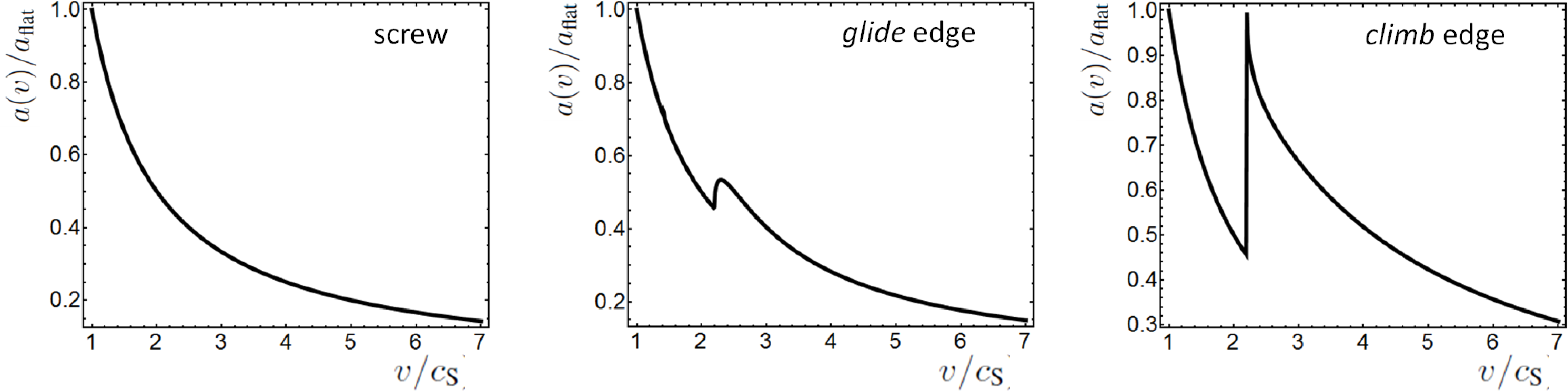}
\end{center}
\caption{\label{fig:fig2} Effective width $a(v)$ as a function of velocity $v$.}
\end{figure}
However, it is possible to adjust empirocally the width of the regularized dislocation as a function of the velocity to bring the asymptotic values of the forces on the regularized and on the flat-core dislocations into exact coincidence, using the width $a=a_{\text{flat}}$ of the flat-core dislocation as a fixed reference. This velocity-dependent `effective width' is shown in Fig.\ \ref{fig:fig2}. Contracting the effective width increases the drag, as this brings the stress field closer to its singular Volterra limit. Such an empirical velocity-dependence is independent from the intrinsic steady-state velocity dependence of $a_{\text{flat}}$ \citep{ROSA01,PELL14}, which is ignored in the present work.

\section{Concluding discussion}
\label{sec:concl}
In an effort to better understand the effect of regularization procedures on radiative drag and inertia, the elastodynamic Peach-Koehler self-force has been computed analytically in isotropic continuum elasticity for a fully-regularized straight dislocation, for prescribed, velocity-independent, core shape. Equations \eqref{eq:partm}, \eqref{eq:massform}, and \eqref{eq:impulsionform} are the main result of the paper. Within the model considered they hold for motion at any velocities and accelerations. They could be generalized to anisotropic materials.

The calculation proceeds almost straightforwardly from a regularized version of Mura's Green-operator expression of the dynamic stress field and Rogula's definition of the self-force. It requires the `standard' energy-related velocity-dependent functions (i.e., mass, impulsion, Lagrangian) to be separated into wave-specific parts. A key step consists in making each partial mass function well-behaved at zero speed, to ensure the vanishing of a boundary term in one partial integration. In the same manner as for the flat-core (Peierls-Eshelby) dislocation, the self-force can be expressed in either `mass' or 'impulsion' forms, namely, as an integral over past times whose integrand involves partial-wave mass or impulsion functions. A closed-form analytical expression for the mass functions is derived in terms of Green-operator components, Eq.\ \eqref{eq:partm}. In the Volterra limit, these regularized expression go to the ill-defined expression \eqref{eq:template}, with no `undetermined' term, and the partial mass functions combine additively into the 'standard' one.

The present derivation, which unambiguously defines the mass functions as second derivatives of Lagrangian functions, rules out earlier alternative definitions of the mass, such as by \cite{SAKA91}. It moreover makes it quite plausible that, irrespective of the regularization procedure, the need for partial-wave decompositions of all energy-related functions should be the rule rather than the exception, although confirmation would require trying out alternative regularization procedures. In this respect, the flat-core dislocation model, which only needs the `standard' energy-related functions, but whose dynamic force involves a separate radiation term outside the time integral, stands as a very peculiar one.

From the physical standpoint, regularization is shown to have a profound effect on the steady-state limit of the self-force, which is an important conclusion in the perspective of predicting high-velocity stress-strain mobility laws for dislocations. Computed with same fixed core width, the radiative drags of the regularized and flat-core dislocations are markedly different at high velocities. However, they can be brought into coincidence in the steady state by introducing an empirical velocity-dependent effective core width $a(v)$ for the regularized model, which suggests that the regularization considered might be quite versatile. However, time-dependence of the core width has not been considered, which is left to future work.\footnote{To account for it, the quantity $a=2\varepsilon$ should be replaced by $[a(t)+a(t')]/2$ in Eq.\ \eqref{eq:fpk}, while the definition of the dislocation current should be re-examined.}

Finally, it is recalled that that besides the PK term, the dynamic force on a dislocation also comprises a contribution akin to the Lorentz force in electrodynamics \cite{NABA51b,ESHE56b,MURA87}. \citet{GURR20} consider it a `mathematical curiosity' as it operates orthogonally to the velocity vector, does no work \citep{LUND96,LUND98}, and cannot play any role if the dislocation is constrained to a unique slip plane, or if a loop translates with uniform velocity \citep{MALE70b}. Its expression reads \citep{ROGU65}
\begin{align}
f_i^{\text{L}}(t)&=-\rho\int\dd^3 r\,v_k(\br,t)I_{ki}(\br,t).
\end{align}
It can be easily evaluated by the method of Sec.\ \ref{sec:straightreg}. Due to the vanishing of the Green operator's gradient components $G_{11,2}$, $G_{12,1}$, $G_{22,2}$, and $G_{33,2}$, on which it exclusively depends, the result turns out to be identically zero for a straight dislocation, whatever the character. However, Mal\'en's analysis (see above reference) suggests that the case of a loop with non-uniform local velocities should deserve a detailed investigation.

\providecommand{\natexlab}[1]{#1}
\newcommand{\enquote}[1]{``#1''}

\appendix
\section{Elements of regularized Green function}
\label{sec:elements}
The nonzero Green function elements of interest read, with $\zS=\vS/\cS$ and $\zL=\vL/\cL$,
\begin{subequations}
\begin{align}
\label{eq:dtg11}
\dot{G}^{(2)a}_{11}(\xi(t)-\xi(t'),0,t-t')
&=\frac{\cS^2}{2\pi\mu}\Re\frac{1}{[\xi(t)-\xi(t')]^2}
\left[\frac{1-2\,\zL^2}{\left(1-\zL^2\right)^{3/2}}-\frac{1}{\sqrt{1-\zS^2}}\right],\\
\label{eq:dtg22}
\dot{G}^{(2)a}_{22}(\xi(t)-\xi(t'),0,t-t')
&=\frac{\cS^2}{2\pi\mu}\Re\frac{1}{[\xi(t)-\xi(t')]^2}
\left[\frac{1-2\,\zS^2}{\left(1-\zS^2\right)^{3/2}}-\frac{1}{\sqrt{1-\zL^2}}\right],\\
\label{eq:dtg33}
\dot{G}^{(2)a}_{33}(\xi(t)-\xi(t'),0,t-t')
&=-\frac{\cS^2}{2\pi\mu}\Re\frac{1}{[\xi(t)-\xi(t')]^2}\frac{\zS^2}{\left(1-\zS^2\right)^{3/2}},\\
\label{eq:g111}
G^{(2)a}_{11,1}(\xi(t)-\xi(t'),0,t-t')
&=\frac{\cS}{2\pi\mu}\Re\frac{1}{[\xi(t)-\xi(t')]^2}
\left[\frac{2-\zS^2}{\zS\sqrt{1-\zS^2}}-\frac{\cS}{\cL}\frac{2-3\,\zL^2}{\zL\left(1-\zL^2\right)^{3/2}}\right],\\
\label{eq:g122}
G^{(2)a}_{12,2}(\xi(t)-\xi(t'),0,t-t')
&=\frac{\cS}{2\pi\mu}\Re\frac{1}{[\xi(t)-\xi(t')]^2}
\left[\frac{\cS}{\cL}\frac{2-\zL^2}{\zL\sqrt{1-\zL^2}}-\frac{2-\zS^2}{\zS\sqrt{1-\zS^2}}\right],\\
\label{eq:g221}
G^{(2)a}_{22,1}(\xi(t)-\xi(t'),0,t-t')
&=\frac{\cS}{2\pi\mu}\Re\frac{1}{[\xi(t)-\xi(t')]^2}
\left[\frac{\cS}{\cL}\frac{2-\zL^2}{\zL\sqrt{1-\zL^2}}-\frac{2-3\,\zS^2}{\zS\left(1-\zS^2\right)^{3/2}}\right],\\
\label{eq:g331}
G^{(2)a}_{33,1}(\xi(t)-\xi(t'),0,t-t')
&=\frac{\cS}{2\pi\mu}\Re\frac{1}{[\xi(t)-\xi(t')]^2}
\frac{\zS^3}{\left(1-\zS^2\right)^{3/2}}.
\end{align}
\end{subequations}
From Eqs.\ \eqref{eq:dtg11}--\eqref{eq:g331}, and definitions \eqref{eq:dtHdef}, \eqref{eq:gradHdef}, one deduces after subtraction of leading-order terms as discussed in the main text, that
\begin{subequations}
\begin{align}
\label{eq:H11}
H^{\rm S}_{11}(z)&=-(1-z^2)^{-1/2}+1,\qquad
H^{\rm L}_{11}(z)=(1-2\,z^2)(1-z^2)^{-3/2}-1,\\
\label{eq:H22}
H^{\rm S}_{22}(z)&=(1-2\,z^2)(1-z^2)^{-3/2}-1,\qquad
H^{\rm L}_{22}(z)=-(1-z^2)^{-1/2}+1,\\
\label{eq:H33}
H^{\rm S}_{33}(z)&=-z^2(1-z^2)^{-3/2},\qquad
H^{\rm L}_{33}(z)=0,\\
\label{eq:H111}
H^{\rm S}_{111}(z)&=z^{-1}(2-z^2)(1-z^2)^{-1/2}-2z^{-1},\qquad
H^{\rm L}_{111}(z)=-(\phi\,z)^{-1}(2-3\,z^2)(1-z^2)^{-3/2}+2(\phi\,z)^{-1},\\
\label{eq:H122}
H_{122}^{\rm S}(z)&=-z^{-1}(2-z^2)(1-z^2)^{-1/2}+2 z^{-1},\qquad
H_{122}^{\rm L}(z)=(\phi\,z)^{-1}(2-z^2)(1-z^2)^{-1/2}-2 (\phi\,z)^{-1},\\
\label{eq:H221}
H_{221}^{\rm S}(z)&=-z^{-1}(2-3\,z^2)(1-z^2)^{-3/2}+2 z^{-1},\qquad
H_{221}^{\rm L}(z)=(\phi\,z)^{-1}(2-z^2)(1-z^2)^{-1/2}-2 (\phi\,z)^{-1},\\
\label{eq:H331}
H_{331}^{\rm S}(z)&=z^3(1-z^2)^{-3/2},\qquad
H_{331}^{\rm L}(z)=0.
\end{align}
\end{subequations}

\end{document}